\documentclass[10pt,twocolumn]{ieeeconf}
\usepackage{color}
\usepackage{float}
\usepackage{amsmath}
\usepackage{amssymb,eqnarray}
\usepackage{mathabx}
\usepackage{graphicx}
\usepackage{stmaryrd}
\usepackage{psfrag}
\usepackage[noend]{algorithmic}
\usepackage{algorithm2e}

\makeatletter
\newcommand{\pushright}[1]{\ifmeasuring@#1\else\omit\hfill$\equationstyle#1$\fi\ignorespaces}
\newcommand{\pushleft}[1]{\ifmeasuring@#1\else\omit$\equationstyle#1$\hfill\fi\ignorespaces}
\makeatother

\usepackage[noend]{algorithmic}
\usepackage{algorithm2e}
\usepackage[english]{babel}
\usepackage{subfig}
\usepackage[export]{adjustbox}
\usepackage{cite}
\usepackage{mathtools}
\mathtoolsset{showonlyrefs=false}

\floatstyle{ruled}
\newfloat{algorithm}{tbp}{loa}
\providecommand{\algorithmname}{Algorithm}
\floatname{algorithm}{\protect\algorithmname}

\newtheorem{thm}{\protect\theoremname}
\newtheorem{defi}{\protect\definitionname}
\newtheorem{prop}{\protect\propositionname}

\newtheorem{problem}{\protect\probname}

\providecommand{\definitionname}{\textbf{Definition}}
\providecommand{\propositionname}{\textbf{Proposition}}
\providecommand{\remarkname}{\textbf{Remark}}
\providecommand{\theoremname}{\textbf{Theorem}}
\providecommand{\lemmaname}{Lemma}
\providecommand{\assumname}{\textbf{Assumption}}
\providecommand{\probname}{\textbf{Problem}}

\IEEEoverridecommandlockouts

\usepackage[format=plain,font=footnotesize,labelfont=bf,labelsep=period]{caption}

\title{Barrier Functions for Multiagent-POMDPs with DTL Specifications}
\author{Mohamadreza Ahmadi, Andrew Singletary, Joel W. Burdick, and Aaron D. Ames \thanks{This work was supported by DARPA Subterranean Challenge. The authors are with the California Institute of Technology, 1200 E. California Blvd., MC 104-44, Pasadena, CA 91125,  e-mail: (\{mrahmadi,asingletary, jwb,ames\}@caltech.edu).)}}

\begin{document}

\maketitle

\begin{abstract}
Multi-agent partially observable Markov decision processes (MPOMDPs) provide a framework to represent heterogeneous autonomous agents subject to uncertainty and partial observation. In this paper, given a nominal policy provided by a human operator or a conventional planning method, we propose a technique based on barrier functions to design a minimally interfering \emph{safety-shield} ensuring satisfaction of  high-level specifications in terms of linear distribution temporal logic (LDTL).  To this end, we use  sufficient and necessary conditions for the invariance of a given set based on discrete-time barrier functions (DTBFs) and formulate sufficient conditions for finite time DTBF to study finite time convergence to a set. We then show that different LDTL mission/safety specifications can be cast as a set of invariance or  finite time reachability problems. We demonstrate that the proposed method for safety-shield synthesis  can be implemented online by a sequence of one-step greedy algorithms. We demonstrate the efficacy of the proposed  method using experiments involving a team of~robots.
\end{abstract}

\vspace{-0.4cm}

\section{Introduction}

Decision making under uncertainty and partial observation is an important branch of artificial intelligence (AI) and probabilistic robotics that has received attention is the recent years.  The applications run the gamut of autonomous driving~\cite{hubmann2018automated} to Mars rover navigation~\cite{nilsson2018toward}.

A popular formalism that can capture the decision making, uncertainty, and partial observation associated with such systems is the partially observable Markov decision process (POMDP). In a POMDP framework, an autonomous agent is not aware of the exact state of the environment and, through a sequence of actions and observations, it updates its \emph{belief} in the current state of the environment. Decision making is then carried out based on the history of the observations or the current belief. Despite the fact that POMDPs provide a unique modeling paradigm, they are  notoriously hard to solve. In particular, it was shown that the infinite-horizon total
undiscounted/discounted/average reward problem for a single agent is undecidable~\cite{madani1999undecidability} and even the finite-horizon  problem for multiple agents with full communication  is  PSPACE-complete~\cite{papadimitriou1987complexity}. However, methods based on discretization of the belief space (known as point-based methods)~\cite{shani2013survey}, heuristics~\cite{amato2015scalable,smith2004heuristic}, finite-state controllers~\cite{sharan2014finite}, or abstractions~\cite{haesaert2018temporal} are shown to be successful to handle relatively large problems. 

\begin{figure}[t] \centering{
\includegraphics[scale=.25]{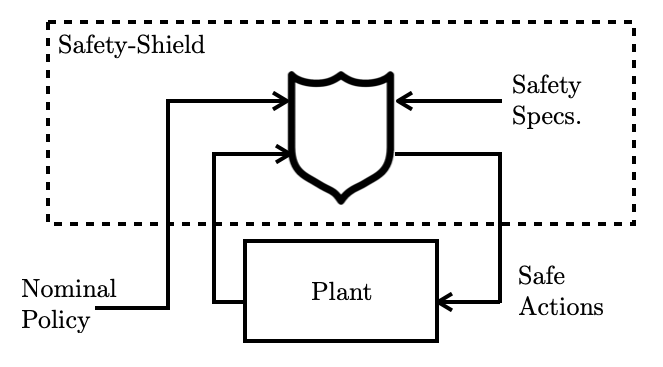}
\caption{Schematic diagram of the proposed \textit{Safety-Shield} framework.}} \label{fig:marssurface}
\vspace{-.5cm}
 \end{figure}

However, for safety-critical systems, such as Mars rovers and autonomous vehicles, safe operation is as (if not more) important than optimality and it is often cumbersome to design a policy to guarantee both safety and optimality. In particular for high-level safety specifications in terms of temporal logic, synthesizing a policy satisfying the specification is undecidable~\cite{chatterjee2016decidable} and  requires heuristics and ad-hoc methods~\cite{chatterjee2015qualitative}. Run-time enforcement of linear temporal logic (LTL) specifications in the absence of partial observability, \textit{i.e.}, in Markov decision processes (MDPs), was considered in~\cite{alshiekh2018safe,jansen2018shielded,hasanbeig2019reinforcement}, where the authors use  automaton representations of the LTL specifications. Though effective for MDPs, the latter approach is not applicable to POMDPs for two main reasons. First, given an LTL formula, the automaton representation can introduce finite but arbitrary large number of additional states. Therefore, model checking of even abstraction-based methods~\cite{wintererabstraction} for POMDPs may not be suitable for run-time enforcement. Second, LTL may not be a suitable logic for describing safety specifications for systems subject to unavoidable uncertainty and partial observation~\cite{jones2013distribution}. Therefore, we use LDTL~\cite{jones2013distribution,vasile2016control}, which can be used for specifying tasks for stochastic systems with partial state information.

In this paper, instead of automaton representations, we employ discrete time barrier functions (DTBFs) to enforce safety/mission specifications in terms of LDTL  specifications in Multi-agent POMDPs (plant) in \emph{run time} with \emph{minimum interference} (see Fig. 1). To this end, we begin by representing the joint belief evolution of an MPOMDP as a discrete-time system~\cite{ahmadi2019control}. We then formulate DTBFs to enforce invariance and finite time reachability and propose Boolean compositions of DTBFs. We propose a safety-shield method based on one-step greedy algorithms to  synthesize a safety-shield for an MPOMDP given a nominal planning policy. We illustrate the efficacy of the proposed approach by applying it to an exploration scenario of a team of heterogeneous robots in  ROS simulation environment. 

The rest of the paper is organized as follows. The next section reviews some preliminary notions and definitions used in the sequel. In Section~\ref{sec:DTBFs}, we propose DTBFs for invariance and finite-time reachability as well as their Boolean compositions. In Section~\ref{sec:safempomdp}, we design a safety-shield based on DTBFs for LDTL specifications. In Section~\ref{sec:example}, we elucidate our results with a multi-robot case study. Finally, in Section~\ref{sec:conclusions}, we conclude the paper.

\vspace{0.1cm}
\noindent\textbf{Notation:}
$\mathbb{R}^n$ denotes the $n$-dimensional Euclidean space. $\mathbb{R}_{\ge 0}$ denotes the set $[0,\infty)$. $\mathbb{N}_{\ge0}$ denotes the set of non-negative integers. For a finite-set $A$,  $|A|$ and $2^A$ denote the number of elements in $A$ and the power set of $A$, respectively. A continuous function $\alpha:[0,a)\to \mathbb{R}_{\ge 0}$ is a class $\mathcal{K}$ function if $\alpha(0)=0$ and it is strictly increasing. Similarly,  a continuous function $\beta:[0,a)\times \mathbb{R}_{\ge 0} \to \mathbb{R}_{\ge 0}$ is a class $\mathcal{KL}$ function if $\beta(r,\cdot) \in \mathcal{K}$ and if  $\beta(\cdot,s)$ is decreasing with respect to $s$ and  $\lim_{s \to \infty} \beta(\cdot,s) \to 0$. For two functions $f:\mathcal{G}\to \mathcal{F}$ and $g:\mathcal{X}\to \mathcal{G}$, $f \circ g:\mathcal{X} \to \mathcal{F}$ denotes the composition of $f$ and $g$ and $\mathrm{Id}:\mathcal{F} \to \mathcal{F}$ denotes the identity function satisfying $\mathrm{Id}\circ f = f$ for all functions~$f:\mathcal{X} \to \mathcal{F}$. The Boolean operators are denote by $\neg$ (negation), $\lor$ (conjunction), and $\land$ (disjunction). The temporal operators are denoted by $\bigcirc$ (next), $\mathfrak{U}$ (until), $\Box$ (always), and $\diamondsuit$ (eventuality). 
\vspace{-.2cm}
\section{Preliminaries}

In this section, we briefly review some notions and definitions used throughout the paper.
\vspace{-.5cm}
\subsection{Multi-Agent POMDPs}

 An MPOMDP~\cite{messias2011efficient,amato2015scalable} provides a sequential  decision-making formalism for high-level planning of multiple autonomous agents under partial observation and uncertainty. At every time step, the agents take actions and receive observations. These observations are shared via (noise and delay free) communication and the agents decide in a centralized framework.
\vspace{0.2cm}
\begin{defi}\label{defi:mpomdp}
\textit{
An MPOMDP is a tuple $\left(I, Q,p^0, \{A_i\}_{i\in I},T,R,\{Z_i\}_{i \in I},O\right)$, wherein 
	\begin{itemize}
        \item $I$ denotes a index set of agents;
		\item $Q$ is a finite set of states with indices $\{1,2,\ldots,n\}$;
		\item $p^0:Q\rightarrow[0,1]$ defines the initial state distribution;
		\item $A_i$ is a finite set of actions for agent $i$ and $A=\times_{i\in I}A_i$ is the set of joint actions;
		\item $T:Q\times A\times Q\rightarrow [0,1]$ is the transition probability, where $
		T(q,a,q'):=P(q^t=q'|q^{t-1}=q,a^{t-1}=a),~\\
		\forall t\in\mathbb{Z}_{\ge 1}, q,q'\in Q, a\in A, $
i.e., the probability of moving to state $q'$ from $q$ when the joint actions $a$ are taken;
\item $R:Q \times A\rightarrow \mathbb{R}$	is the immediate reward function for taking the joint action $a$ at state $q$;
    		\item $Z_i$ is the set of all possible observations for agent $i$ and $Z=\times_{i \in I} Z_i$, representing outputs of  discrete sensors, e.g. $z\in Z$ are incomplete projections of the world states $q$, contaminated by sensor noise;
		\item $O:Q\times A \times Z\rightarrow [0,1]$ is the observation probability (sensor model), where $
		O(q',a,z):=P(z^t=z|q^{t}=q',a^{t-1}=a),~
	    \forall t\in\mathbb{Z}_{\ge 1}, q\in Q, a\in A, z\in Z, $
i.e., the probability of seeing joint observations $z$ given joint actions $a$ were taken and resulting in state $q'$.
	\end{itemize}
	}
\end{defi}
\vspace{0.2cm}
 	Since the states are not directly accessible in an MPOMDP, decision making requires the history of joint actions and joint observations. Therefore, we must define the notion of a joint \emph{belief} or the posterior as sufficient statistics for the history. Given an MPOMDP, the joint belief at $t=0$ is defined as $b^0(q)=p^0(q)$ and $b^t(q)$ denotes the probability of the system being in state $q$ at time $t$. At time $t+1$, when joint action $a\in A$ is taken and joint observation $z \in Z$ is observed, the belief is updated via a Bayesian filter as
\begin{align} 
& b^t(q') \label{equation:belief update}
:= BU(z^t,a^{t-1},b^{t-1}) \nonumber\\
&= \frac{O(q',a^{t-1},z^{t})\sum_{q\in Q}T(q,a^{t-1},q')b^{t-1}(q)}{\sum_{q'\in Q}O(q',a^{t-1},z^{t})\sum_{q\in Q}T(q,a^{t-1},q')b^{t-1}(q)}
\end{align}

where the beliefs belong to the belief unit simplex
$$
\mathcal{B} = \left\{ b \in [0,1]^{|Q|} \mid \sum_{q\in Q} b^t(q)=1,~\forall t  \right\}.
$$

A policy in an MPOMDP setting is then a mapping $\pi:\mathcal{B} \to A$, i.e., a mapping from the continuous joint beliefs space into the discrete and finite joint action space. When $I$ is just a singleton (only one agent), we have a POMDP~\cite{smallwood1973optimal}.


\vspace{-.2cm}
\subsection{Linear Distribution Temporal Logic}

We formally describe high-level mission specifications that
are defined in temporal logic. Temporal logic has been used
as a formal way to allow the user to  intuitively
specify high-level specifications, in for example, robotics~\cite{lahijanian2010motion}. The temporal logic we use in this paper can be used for specifying tasks for stochastic systems with partial
state information. This logic is suitable for problems involving significant state uncertainty, in which
the state is estimated on-line. The syntactically co-safe linear distribution temporal logic (scLDTL) describes co-safe linear temporal logic properties of probabilistic systems~\cite{jones2013distribution,vasile2016control}. We consider a modified version to scLDTL, the linear  distribution temporal logic (LDTL), which includes the additional temporal operator $\Box$ ``always''. The latter operator is important since it can be used to describe notions such as \emph{safety}, \emph{liveness}, and \emph{invariance}.

LDTL has predicates of the type $\zeta<0$ with $\zeta \in F_Q=\{f \mid f: \mathcal{B} \to \mathbb{R}\}$, and state predicates $q \in A$ with $A \in 2^Q$. 
\vspace{0.1cm}
\begin{defi}[LDTL Syntax]\textit{
An LDTL formula over predicates $F_Q$ and $Q$ is inductively defined as 
\begin{equation}
    \varphi := A | \neg A | \zeta | \neg \zeta | \varphi \lor \varphi | \varphi \land \varphi | \varphi~\mathfrak{U}~\varphi | \bigcirc \varphi | \diamondsuit \varphi | \Box \varphi,
    \end{equation}
    where $A \in 2^Q$ is a set of states, $\zeta \in F_Q$ is a belief predicate, and $\varphi$ is an LDTL formula.}
\end{defi}

Satisfaction over pairs of hidden state paths and sequences of belief states can then be defined as follows.

\vspace{0.1cm}
\begin{defi}[LDTL Semantics]\label{defi:scLDTLsemantics}
\textit{The semantic of LDTL formulae is defined over words $\omega \in (Q \times \mathcal{B})^\infty$. Let $(q^i,b^i)$ be the $i$th letter in $\omega$. The satisfaction of a LDTL formula $\varphi$ at position $i$ in $\omega$, denoted by $\omega^i \models \varphi$ is recursively defined as follows
\begin{itemize}
    \item  $\omega^i \models A$ if $q^i \in A$,
    \item $\omega^i \models \neg A$ if $q^i \notin A$,
    \item $\omega^i \models f$ if $f(b^i) <0$,
    \item $\omega^i \models \neg f$ if $f(b^i) \ge0$,
    \item $\omega^i \models \varphi_1 \land \varphi_2$ if $\omega^i \models \varphi_1 $ and $\omega^i \models  \varphi_2$,
    \item $\omega^i \models \varphi_1 \lor \varphi_2$ if $\omega^i \models \varphi_1 $ or $\omega^i \models  \varphi_2$,
    \item $\omega^i \models \bigcirc \varphi$ if $\omega^{i+1} \models \varphi$,
    \item $\omega^i \models \varphi_1 \mathfrak{U} \varphi_2$ if there exists a $j \ge i$ such that $\omega^j \models \varphi_2$ and for all $i \le k < j$ it holds that $\omega^k \models \varphi_1$, 
    \item $\omega^i \models \diamondsuit \varphi$ if there exists $j \ge i$ such that $\omega^j \models \varphi$, and
    \item $\omega^i \models \Box \varphi$ if, for all $j \ge i$,  $\omega^j \models \varphi$.
\end{itemize}
The word $\omega$ satisfies a formula $\varphi$, i.e., $\omega \models \varphi$, iff $\omega^0 \models \varphi$.
}
\end{defi}
\vspace{-.3cm}


\section{Discrete-Time Barrier Functions}\label{sec:DTBFs}

In order to guarantee the satisfaction of  the LDTL formulae in MPOMDPs, we use barrier functions~\cite{ames2017control} rather than automatons and model checking. Hence, our method  does not rely on automaton representations,  discretizations  of the belief space, or finite memory controllers. These barrier functions ensure that the solutions to the joint belief update equation remain inside or reach subsets of the belief simplex that is induced by the LDTL formula. Noting that the joint belief evolution of an MPOMDP~(1) can be described by a discrete-time system~\cite{ahmadi2019control}, in this  section, we propose conditions based on DTBFs for verifying invariance and finite-time reachability properties.

Given an MPOMDP as defined in Definition~1, the joint belief update equation~(1) can be described by the following discrete-time system
\begin{equation}\label{eq:discdyn}
b^{t+1} = f(b^{t}),~~t \in \mathbb{N}_{\ge 0},
\end{equation}
with $f:\mathcal{B}  \to \mathcal{B} \subset \mathbb{R}^n$ given an observation and an action. We consider subsets of the belief simplex defined as
\begin{subequations}\label{eq:safeset}
\begin{eqnarray}
\mathcal{S} :=\{ b \in \mathcal{B} \mid h(b) \ge 0 \}, \\
\mathrm{Int}(\mathcal{B}) :=\{ b \in \mathcal{B} \mid h(b) > 0 \}, \\
\partial \mathcal{S} :=\{ b \in \mathcal{B} \mid h(b) = 0 \}.
\end{eqnarray}
\end{subequations}
\vspace{0.1cm}
We then have the following definition of a DTBF.

\begin{defi}[Discrete-Time Barrier Function]\label{def:dtbf}
\textit{
For~the discrete-time system~\eqref{eq:discdyn}, the continuous function $h : \mathbb{R}^n \to \mathbb{R}$ is a discrete-time barrier
function for the set $\mathcal{S}$ as defined in~\eqref{eq:safeset}, if there exists $\alpha \in \mathcal{K}$ satisfying $\alpha(r) < r$ for all $r>0$  such that
\begin{equation}\label{eq:BFinequality}
    h(b^{t+1})-h(b^{t}) \ge - \alpha(h(b^{t})),\quad \forall b \in \mathcal{B}.
    \end{equation}
    }
    \end{defi}
\vspace{0.1cm}

 In fact, the DTBF defined above is a discrete-time zeroing barrier function per the literature \cite{ames2017control} (see also the reciprocal DTBF proposed in~\cite{agrawal2017discrete}), but we drop the ``zeroing'' as it is the only form of barrier function that will be considered in this paper.

We can show that the existence of a DTBF is both necessary and sufficient for invariance. We later show in Section V that such DTBF can be used to verify a class of LDTL specifications.


\vspace{0.1cm}
\begin{thm}[\!\!\cite{2019arXiv190307823A}]\label{thm:BFdiscrete}
\textit{
Consider the discrete-time system~\eqref{eq:discdyn}. Let $\mathcal{S} \subseteq \mathcal{B} \subset \mathbb{R}^n$ with $\mathcal{S}$ as described in~\eqref{eq:safeset}. Then, $\mathcal{S}$ is invariant if and only if there exists a DTBF as defined in Definition~\ref{def:dtbf}.
}
\end{thm}
\vspace{-0.2cm}

\subsection{Finite Time DTBFs}

Another class of problems we are interested in involve checking whether the solution of a discrete time system can reach a set in finite time. We will show in Section~\ref{sec:safempomdp} that such problems arise when dealing with ``eventuality'' type LDTL specifications. To this end, we define a finite time DTBF (see~\cite{8619113} for the continuous time variant).
\vspace{0.1cm}
\begin{defi}[Finite Time DTBF] \label{def:ftdtbf}
\textit{
For the discrete-time system~\eqref{eq:discdyn}, the  continuous function $\tilde{h}:\mathcal{B} \to \mathbb{R}$ is a finite time DTBF for the set $\mathcal{S}$ as defined in~\eqref{eq:safeset}, if there exist constants  $0<\rho<1$ and $\varepsilon > 0$ such that
\begin{equation}\label{eq:BFft}
    \tilde{h}(b^{t+1})-\rho \tilde{h}(b^{t}) \ge \varepsilon (1-\rho),\quad \forall b \in \mathcal{B}.
\end{equation}
}
\end{defi}
\vspace{0.1cm}

We then  have the following result to check finite time reachability of a set for a discrete-time system. 
\vspace{0.1cm}
\begin{thm} \label{thm:FTDTBF}
\textit{
Consider the discrete-time system~\eqref{eq:discdyn}. Let $\mathcal{S} \subset \mathcal{B} \subset \mathbb{R}^n$ be as described in~\eqref{eq:safeset}. If there exists a finite time DTBF $\tilde{h}$ as in Definition~\ref{def:ftdtbf}, then for all $b^0 \in \mathcal{B}\setminus \mathcal{S}$, there exists a $t^* \in \mathbb{N}_{\ge 0}$ such that $b^{t^*} \in \mathcal{S}$. Furthermore, 
\begin{equation}
    t^* \le {\log\left(\frac{\varepsilon - \tilde{h}(b^0)}{\varepsilon}\right)}/{\log\left(\frac{1}{\rho}\right)},
    \end{equation}
    where the constants $\rho$ and $\varepsilon$ are as defined in Definition~\ref{def:ftdtbf}.
}
\end{thm}
\vspace{0.1cm}
\begin{proof}
We prove by induction. With some manipulation inequality~\eqref{eq:BFft} can be modified to
$$
\tilde{h}(b^{t+1})-\varepsilon \ge \rho \tilde{h}(b^t) - \rho \varepsilon = \rho \left(  \tilde{h}(b^t) -    \varepsilon      \right).
$$
Thus, for $t=0$, we have  
$
\tilde{h}(b^{1})-\varepsilon \ge \rho \left(  \tilde{h}(b^0) -    \varepsilon      \right).
$ 
For $t=1$, we have 
$$
\tilde{h}(b^{2})-\varepsilon \ge \rho \left(  \tilde{h}(b^1) -    \varepsilon      \right) \ge \rho^2 \left(  \tilde{h}(b^0) -    \varepsilon      \right),
$$
where we used the inequality for $t=0$ to obtain the last inequality above. Then, by induction, we have 
$\label{eq:BFgrowthft}
\tilde{h}(b^t) - \varepsilon \ge \rho^t \left(  \tilde{h}(b^0) -    \varepsilon      \right).
$ 
Multiplying both sides with $-1$ gives
\begin{equation}\label{eq:BFgrowthft2}
\varepsilon - \tilde{h}(b^t)  \le  \rho^t \left( \varepsilon - \tilde{h}(b^0)      \right)
\end{equation}
Since $b^0 \in \mathcal{B}\setminus \mathcal{S}$, i.e., $\tilde{h}(b^0)<0$, $\varepsilon - \tilde{h}(b^0)$ is a positive number. Dividing both sides of~\eqref{eq:BFgrowthft2} with the positive quantity $\varepsilon- \tilde{h}(b^0)$ yields
$$
\frac{\varepsilon - \tilde{h}(b^t)}{\varepsilon - \tilde{h}(b^0)} \le \rho^t.
$$
Taking the logarithm of both sides of the above inequality  gives
$$
\log\left(\frac{\tilde{h}(b^t) - \varepsilon}{\tilde{h}(b^0) -    \varepsilon}\right) \le t\log(\rho),
$$
or equivalently 
$$
-\log\left(\frac{\tilde{h}(b^0) - \varepsilon}{\tilde{h}(b^t) -    \varepsilon}\right) \le -t \log(\frac{1}{\rho}).
$$
Since $0<\rho<1$, $\log(\frac{1}{\rho})$ is a positive number. Dividing both sides of the inequality above with the negative number $-\log(\frac{1}{\rho})$ obtains 
$
  t \le {\log\left(\frac{\varepsilon - \tilde{h}(b^0)}{\varepsilon- \tilde{h}(b^t)}\right)}/{\log\left(\frac{1}{\rho}\right)}.
$ 
Because $t$ is a positive integer, $\varepsilon - \tilde{h}(b^0) \ge \varepsilon - \tilde{h}(b^t)$. That is, $\tilde{h}(b^0) \le \tilde{h}(b^t)$ along the solutions $b^t$ of the discrete-time system~\eqref{eq:discdyn}, which implies that $b^t$ approaches $\mathcal{S}$. Also, by definition, $b^t$ reaches $\mathcal{S}$ at least at the boundary at $t^*$ when $\tilde{h}(b^t)=0$. Substituting $\tilde{h}(b^t)=0$ in the last inequality for $t$ gives  $t^* \le {\log\left(\frac{\varepsilon - \tilde{h}(b^0)}{\varepsilon}\right)}/{\log\left(\frac{1}{\rho}\right)}$, which gives an upper bound for the first time $b^t \in \mathcal{S}$.
\end{proof}



\subsection{Boolean Composition of Finite Time DTBFs}\label{sec:boolean}

In order to assure specifications involving conjunction or disjunction of LDTL formulae in Definition~\ref{defi:scLDTLsemantics}, we need to consider properties of sets defined by Boolean composition of DTBFs. In this regard, in~\cite{glotfelter2017nonsmooth}, the authors proposed non-smooth barrier functions as a means to analyze composition of barrier functions by Boolean logic, i.e., $\lor$, $\land$, and $\neg$. Similarly, in this study,  we propose non-smooth DTBFs. 
The negation operator is trivial and can be shown by checking if $-h$ satisfies the corresponding property.

In the following, we propose conditions for checking Boolean compositions of finite time DTBFs. Fortunately, since we are concerned with discrete time systems, this does not require non-smooth analysis (for a similar result pertaining compositions of DTBFs see Proposition 1 in \cite{2019arXiv190307823A}).

\vspace{0.1cm}
\begin{prop} \label{prop:booleanftBFs}
\textit{
Let $\mathcal{S}_i = \{ b \in \mathcal{B} \mid \tilde{h_i}(b)\ge 0\}$, $i=1,\ldots,k$ denote a family of sets defined analogous to $\mathcal{S}$ in~\eqref{eq:safeset}. Consider the discrete-time system~\eqref{eq:discdyn}. If there exist constants $0<\rho<1$ and $\varepsilon>0$ such that 
\begin{equation}\label{eq:disjftBF1}
\min_{i=1,\ldots,k} \tilde{h_i}(b^{t+1}) - \rho \min_{i=1,\ldots,k} \tilde{h_i}(b^{t}) \ge \varepsilon (1-\rho),~\forall b \in \mathcal{B},
\end{equation}
then there exists 
\begin{equation}\label{adcasasdfasdafasdas}
  t^{*} \le {\log\left(\frac{\varepsilon - \min_{i=1,\ldots,k}\tilde{h_i}(b^0)}{\varepsilon}\right)}/{\log\left(\frac{1}{\rho}\right)}.
  \end{equation}
 such that if $b^0 \in \mathcal{B} \setminus \bigcup_{i=1}^k\mathcal{S}_i$ then $b^{t^*} \in \left \{ b \in \mathcal{B} \mid  \land_{i =1,\ldots,k} \left(\tilde{h_i}(b) \ge 0\right) \right\}$. Similarly, the disjunction case follows by replacing $\min$ with $\max$ in~\eqref{eq:disjftBF1} and~\eqref{adcasasdfasdafasdas}.}
\end{prop}
\vspace{0.1cm}
\begin{proof}
We prove the conjunction case and the disjunction case follows the same lines. If~\eqref{eq:disjftBF1} holds, from the proof of Theorem~2, we can infer that
$$
\min_{i=1,\ldots,k}\tilde{h_i}(b^t) - \varepsilon \ge \rho^t \left(  \min_{i=1,\ldots,k}\tilde{h_i}(b^0) -    \varepsilon      \right),
$$
which implies that 
$
  t \le {\log\left(\frac{\varepsilon - \min_{i=1,\ldots,k}\tilde{h_i}(b^0)}{\varepsilon- \min_{i=1,\ldots,k}\tilde{h_i}(b^t)}\right)}/{\log\left(\frac{1}{\rho}\right)}.
$ 
 If $b^0 \in \mathcal{B} \setminus \bigcup_{i=1}^k\mathcal{S}_i$, then by definition $\tilde{h_i}(b^0)<0,~i=1,\ldots,k$. Hence, $\min_{i=1,\ldots,k}\tilde{h_i}(b^0)<0$. Moreover, because $t$ is a positive integer, \begin{equation*}\varepsilon - \min_{i=1,\ldots,k}\tilde{h_i}(b^0) \ge \varepsilon - \min_{i=1,\ldots,k}\tilde{h_i}(b^t).\end{equation*} That is, $\min_{i=1,\ldots,k}\tilde{h_i}(b^0) \le \min_{i=1,\ldots,k}\tilde{h_i}(b^t)$ along the solutions $b^t$ of the discrete-time system~\eqref{eq:discdyn}. Furthermore, $b^t \in  \left \{ b \in \mathcal{B} \mid  \land_{i =1,\ldots,k} \left(\tilde{h_i}(b) \ge 0\right) \right\}$ whenever $\min_{i=1,\ldots,k}\tilde{h_i}(b^t) \ge 0$. The upper-bound for this $t$ happens when $\min_{i=1,\ldots,k}\tilde{h_i}(b^t) = 0$, i.e., when all $\tilde{h_i}(b^t)$ are either positive or zero. This by definition implies that $b^t \in  \left \{ b \in \mathcal{B} \mid  \land_{i =1,\ldots,k} \left(\tilde{h_i}(b) \ge 0\right) \right\}$. Then, setting $\min_{i=1,\ldots,k}\tilde{h_i}(b^t) = 0$ gives 
 $$
  t^{*} \le {\log\left(\frac{\varepsilon - \min_{i=1,\ldots,k}\tilde{h_i}(b^0)}{\varepsilon}\right)}/{\log\left(\frac{1}{\rho}\right)}.
$$\end{proof}

\section{Safety-Shield Synthesis}\label{sec:safempomdp}

Since the states are not directly observable in MPOMDPs, we are interested in guaranteeing safety specifications in terms of LDTL in a probabilistic setting in the joint belief space.  We denote by $\pi_n : \mathcal{B} \to \mathcal{A}$ a nominal joint policy mapping each joint belief into a joint action. We use subscript $n$ to denote variables corresponding to the nominal policy.

We are interested in solving the following problem.
 \vspace{.1cm}
\begin{problem}[Safety-Shield Synthesis]
\textit{
Given an MPOMDP as defined in Definition~\ref{defi:mpomdp}, a corresponding belief update equation~\eqref{equation:belief update},  a safety LDTL formula $\varphi$, and a nominal planning policy $\pi_n$, determine a sequence of actions $a^t,~t \in \mathbb{N}_{\ge 0}$ such that $\omega^0 = (q^0,b^0)\models \varphi$  and the quantity   $\|r^t - r_n^t\|^2$ is minimized for all $t \in \mathbb{N}_{\ge 0}$, where $r_n^t$ denotes the nominal immediate reward at time step~$t$.
}
\end{problem}
\vspace{-0.4cm}
\subsection{Enforcing LDTL via DTBFs}\label{sec:formDTBF}

\begin{table}[t]
\begin{center}
\begin{tabular}{| c|c|} 
\hline
LDTL Specification & DTBF Implementation  \\
\hline \hline
$\omega^i \models A$ & $h(b^i)= \sum_{q \in A} b^i(q) -1$  \\ 
\hline
$\omega^i \models \neg A$ & $h(b^i)= \sum_{q \in {Q\setminus A}} b^i(q) -1$  \\ 
\hline
$\omega^i \models  f$ & $h(b^i)= -f(b^i)+\delta$\\
\hline
$\omega^i \models  \neg f$ & $h(b^i)= f(b^i)$\\
\hline 
$\omega^i \models  \varphi_1 \land \varphi_2$ & $h(b^i)= \min\{h_1(b^i),h_2(b^i)\}$\\
\hline
$\omega^i \models  \varphi_1 \lor \varphi_2$ & $h(b^i)= \max\{h_1(b^i),h_2(b^i)\}$\\
\hline
$\omega^i \models  \bigcirc \varphi$ & $h(b^{i+1})= h_{\varphi}(b)$\\
\hline
$\omega^i \models   \varphi_1 \mathfrak{U} \varphi_2$ & $h_2(b^j)<0 \implies h=h_1(b^j), \forall j \ge i$\\
\hline 
$\omega^i \models   \diamondsuit \varphi$ & $h(b^j)=\tilde{h}(b^j), ~\forall i \le j \le t^*$\\
\hline
$\omega^i \models   \Box \varphi$ & $h(b^j)=h_{\varphi}(b^j)$, $\forall j \ge i$ \\
\hline
\end{tabular}
\caption{LDTL specifications and the DTBF implementation. \vspace{-1cm}}
\label{table:LDTLvsDTBF}
\end{center}
\end{table}

In this section, we describe how the semantics of LDTL as given in Definition~\ref{defi:scLDTLsemantics} can be represented as set invariance and reachability conditions over the belief simplex. These set invariance/reachability conditions can then be enforced via DTBFs as summarized in Table~\ref{table:LDTLvsDTBF}. We describe each row of the table in more detail below.

Satisfaction of the formulae of the type $\omega^i \models A\subset Q$ can be encoded as verifying whether $q^i \in A$. In the belief simplex, this is equivalent to checking whether $b^i \in \mathcal{B}_s=\{b^i \in \mathcal{B} \mid \sum_{q \in A} b^i(q) \ge 1\}$, which can be checked by considering the DTBF $h(b^i)=\sum_{q \in A} b^i(q)-1$. On the other hand, $\omega^i \models \neg A\subset Q$ can be cast as checking whether $b^i \in \mathcal{B}_s=\{b^i \in \mathcal{B} \mid \sum_{q \in {Q \setminus A}} b^i(q) \ge 1\}$ by considering the DTBF $h(b^i)=\sum_{q \in {Q \setminus A}} b^i(q)-1$. Formulae $\omega^i \models f$ and $\omega^i \models \neg f$ are already defined in the belief space, since $\omega^i \models f$ implies $f(b^i) < 0$ and $\omega^i \models \neg f$ implies $f(b^i) \ge 0$. They can be checked by considering DTBFs $h(b^i)=-f(b^i)+\delta$ with $0<\delta<<1$ for $\omega^i \models f$ and $h(b^i)=f(b^i)$ for $\omega^i \models \neg f$. The Boolean composition of different formulae such as $\omega^i \models \varphi_1 \land \varphi_2$ and $\omega^i \models \varphi_1 \lor \varphi_2$ can be implemented by Boolean composition of the barrier functions as discussed in Section~\ref{sec:boolean}.

Moreover, our proposed safety-shield checks the satisfaction of the LDTL safety properties real-time at every time step. Therefore, it can also verify temporal properties as explained next.  Satisfaction of temporal formulae such as $\omega^i \models \bigcirc \varphi$ can be implemented by checking whether $\varphi$ is satisfied in the next step using DTBFs as discussed above. Specifications of the type $\omega^i \models \varphi_1 \mathfrak{U} \varphi_2$ can be enforced by checking whether formula $\varphi_1$ is satisfied until~$\varphi_2$. To this end, we can check whether formula $\varphi_2$ is not satisfied at every time step $i$ by checking inequality $h_2(b)<0$ where $h_2$ is the DTBF for formula $\varphi_2$. If $\varphi_2$ is not satisfied, then $\varphi_1$ is checked via a corresponding DTBF $h_1$. The specification $\omega^i \models \diamondsuit \varphi$ can be checked using the finite time DTBF given by Theorem~2. Note that the property is checked until $t^*$, since after $t^*$ the formula $\varphi$ is ensured to hold. Hence, the eventually specification is satisfied. The specification $\omega^i \models \Box \varphi$ can be simply checked by checking whether $\varphi$ is satisfied for all time using the corresponding DTBF $h_\varphi$.
\vspace{-.2cm}
\subsection{Safety-Shield Synthesis}

 



 Algorithm 1 illustrates how DTBFs can shield the agent actions to ensure LDTL safety. Note that the choice of whether \eqref{eq:BFinequality} or \eqref{eq:BFft} should be checked at a time step depends on the specification we want to satisfy as shown in  Table~\ref{table:LDTLvsDTBF}.  At every time step $t$, the algorithm first computes the next joint belief $b^t$ given the nominal action $a_n$ designed based on the nominal policy $\pi_n$. It then checks whether that action leads to a safe joint belief update. If yes, the algorithm returns $a_n$ for implementation. If no, the algorithm picks a joint action $a(i)$  from $|A|$ combinations of actions (recall that $\times_{i \in I}A_i =A$). For each joint action $a(i)$, it computes the next joint belief and checks whether  the next joint belief satisfies the safety specification. If the safety specification is satisfied, it computes the corresponding reward function $r(i)$ for the joint action $a(i)$. It then picks a safe joint action that minimally changes the immediate reward from the nominal immediate reward $r_n^t$ in a least squares sense. This ensures that the decision making remains as much faithful as possible to the nominal policy~(see~\cite{gurriet2018online} for analogous formulations for systems described by nonlinear differential equations). 
 
 \begin{algorithm}[t]
    \begin{algorithmic}[1]
        \REQUIRE System information $I$, $Q$, $A$, $T$, $R$, $Z$, $O$, nominal policy $\pi_n$, safety specifications defined by LDTL formula~$\varphi$, current observation $z^t$, the past belief $b^{t-1}$
        
\STATE       $ b^t = BU(z^t,a^{t-1}_n,b^{t-1}) $
\IF{$h(b^{t})-h(b^{t-1}) \ge - \alpha(h(b^{t-1}))$ or   \quad $\tilde{h}(b^{t})-\rho \tilde{h}(b^{t-1}) \ge \varepsilon (1-\rho)$}
 \RETURN $a^* = a_n^t$
 \ELSE
        \STATE  $i=1$ 
         \FOR{$i =1,2,\ldots,|A|$}
\STATE $ b^t = BU(z^t,a(i),b^{t-1})$ \\
  \IF{$h(b^{t})-h(b^{t-1}) \ge - \alpha(h(b^{t-1}))$ or \quad \quad \quad \quad $\tilde{h}(b^{t})-\rho \tilde{h}(b^{t-1}) \ge \varepsilon (1-\rho)$} 
 \STATE $ r(i) = \left(\sum_{q' \in Q} b(q') R(q',a(i))\right)$ 
 \ENDIF
  \ENDFOR
\STATE $i_* =  \arg \min_{i=1,2,\ldots,|A|} \|r(i)-r_n^t\|^2$ 

 \RETURN $a^* = a({i_*})$.
\ENDIF

    \end{algorithmic}
    \caption{The one-step greedy algorithm for safety-shield synthesis given the nominal policy at every time-step $t$.}\label{alg:daclyf}
\end{algorithm}

\section{Case Study: Multi-Robot Exploration} \label{sec:example}

To demonstrate our method, we consider high-fidelity simulations of three heterogeneous robots, namely, a drone and two ground vehicles (a Rover Robotics Flipper and a modified Segway) exploring an unknown environment in ROS. The drone can rapidly explore the environment from above and it is used to locate a desired sample (goal).  The Flipper is a small, tracked vehicle capable of traversing in  rough terrain, whose job is to locate obstacles in the area. The Segway is larger, wheeled robot without external sensing capabilities, whose purpose is to retrieve the sample without colliding any obstacles. For the MPOMDP representation and more details on the setup, we refer the interested reader to Section V in~\cite{2019arXiv190307823A}. 

The mission specification is given by the formula:
\begin{equation}\label{eq:specexample}
    \varphi = \Box \neg (f_1 \lor f_2) \land \diamondsuit f_3,
\end{equation}
where $f_1=0.1-b(q_S)b(q_F)$, $f_2=0.1-\land_{i=1}^3 b(q_S)b(q_{o_i})$, and $f_3=0.5-b(q_S)b(q_G)$. Formula~\eqref{eq:specexample} ensures that the Segway (located at $q_S$) \emph{always}  avoids the Flipper (located at $q_F$) and the three obstacles (located at $q_{o_i}$, $i=1,2,3$) with more than 0.90 probability and \emph{eventually} reaches the goal (located at $q_G$) with more than 0.5 probability.

In order to enforce LDTL formula~\eqref{eq:specexample}, we use the  DTBF $
h(b) = \min(f_1,f_2), $ 
where we used De Morgan's laws to obtain $\neg (f_1 \lor f_2) = \neg f_1 \land \neg f_2$, the fourth row of Table~I, and Proposition~1, and the finite time DTBF
$
\tilde{h}(b)=b(q_S)b(q_G)-0.5
$ 
where we used the third and ninth rows of Table~I.


For the finite time DTBF condition~\eqref{eq:BFft}, the parameters $\rho$ and $\epsilon$ must be set to tune how quickly the set must be reached. To allow for more freedom of operation, we choose $\rho = 0.99$ and $\epsilon = 0.1$.

The nominal policy used for the UAV and the Flipper is a simple implementation of A*, that tries to maximize information gain by moving in new regions of the state space~\cite{drew_uav}. The Segway, on the other hand, is fed a constant action repeatedly, and relies on the safety-shield to reach the sample. 

\begin{figure*}[h]
    \hspace*{0 cm}
     \centering
    \begin{subfloat}
        {\includegraphics[height =0.246\textwidth, valign =t ]{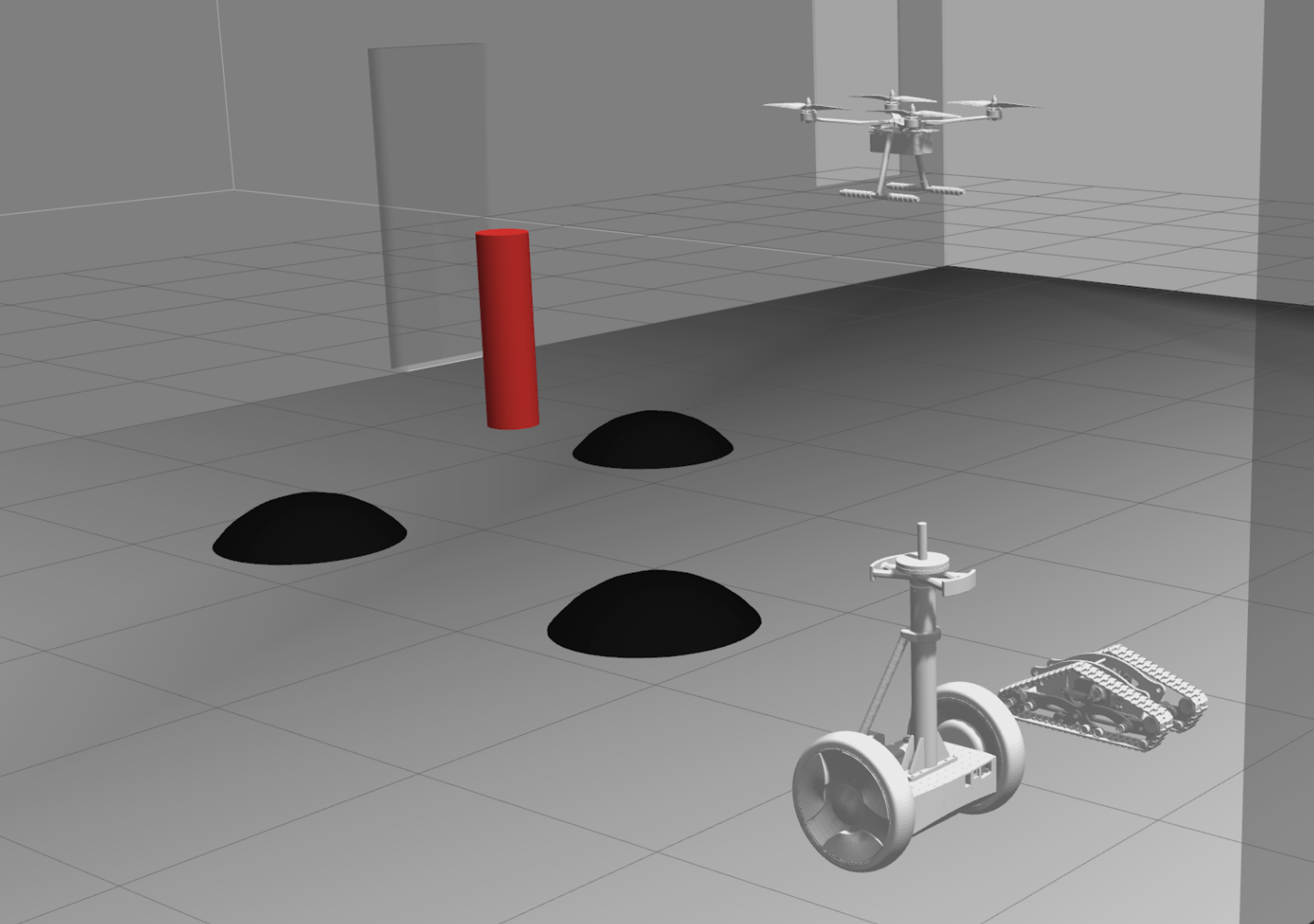}}
    \end{subfloat}
    \hfill
    \hspace*{0 cm}
    \begin{subfloat}
      {\includegraphics[height =0.246\textwidth, valign =t]{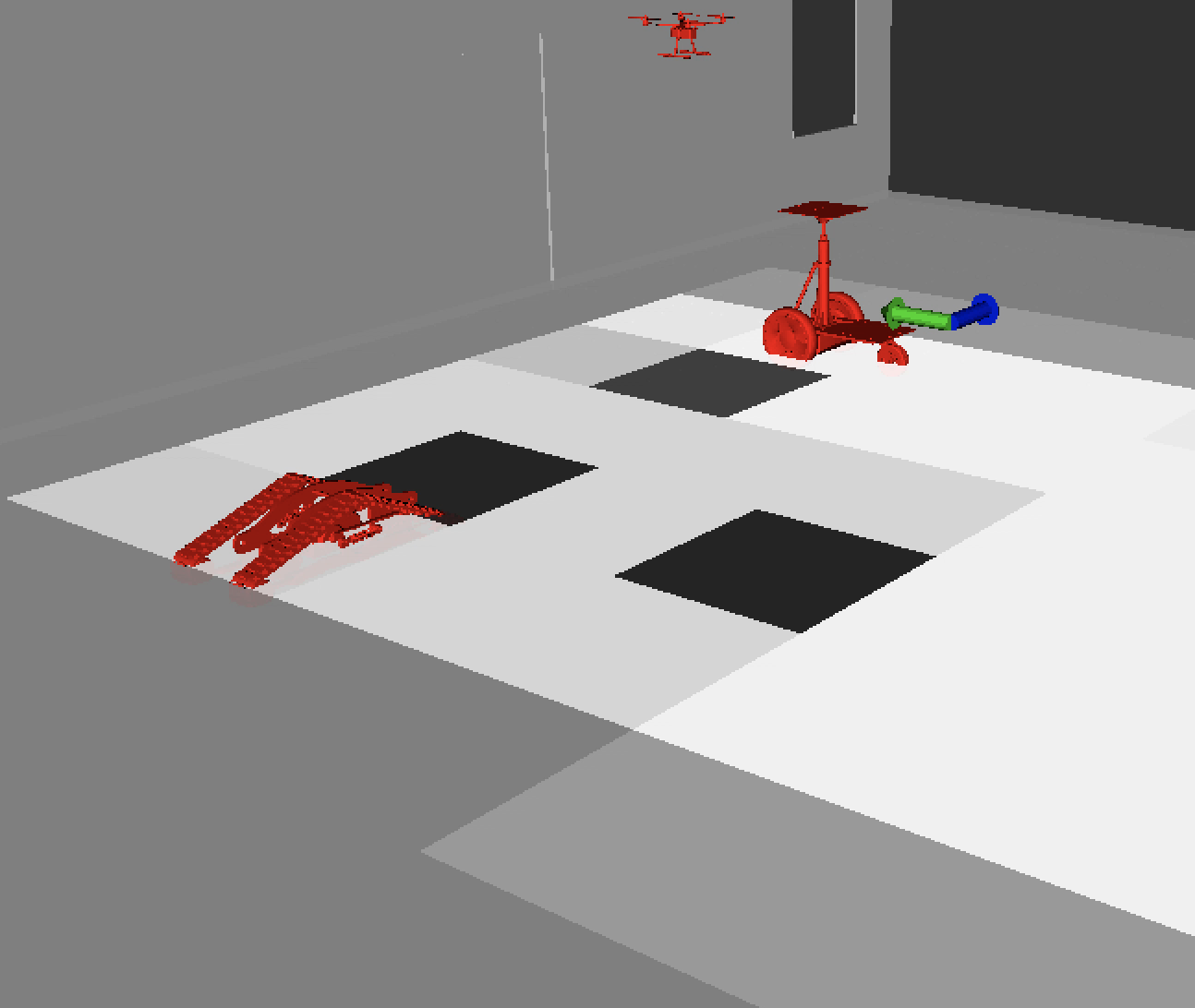}}
    \end{subfloat}
    \hfill
    \begin{subfloat}
     {\includegraphics[height =0.246\textwidth,valign=t]{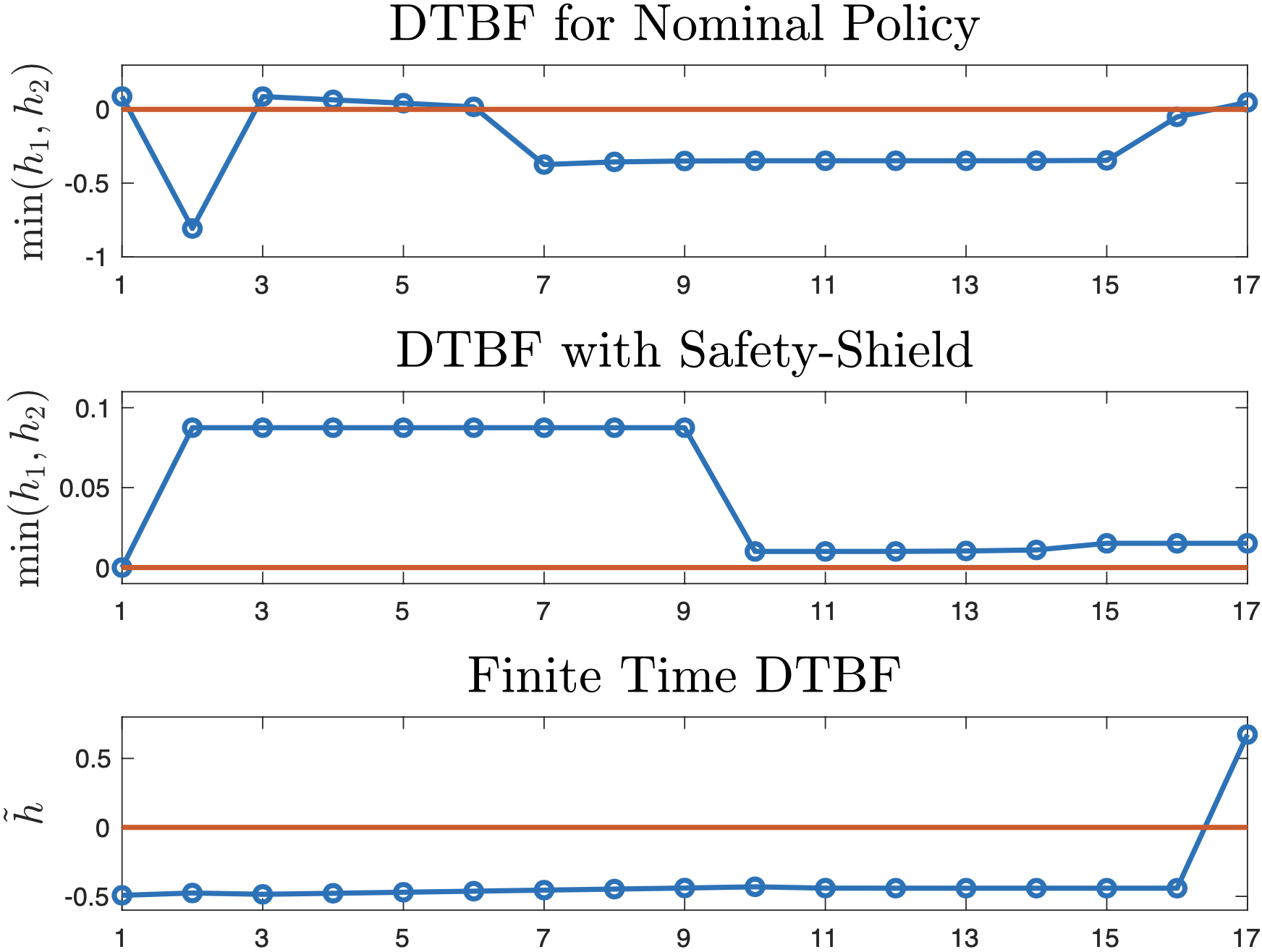}}
    \end{subfloat}
        \caption{Simulation results of the multi-agent system. \textbf{(Left)} The Gazebo simulation environment showcasing the three agents, obstacles (black), and sample (red) at the beginning of the experiment.
        \textbf{(Center)} The RVIZ visualization of the experiment. The costmaps for the obstacles and the sample are overlaid, and the arrows showcase the nominal action (blue) as well as the action given by the safety shield (green).
        \textbf{(Right)} The plots of the DTBFs for the experiment, as explained above.
        }
    \label{fig:sim}
\end{figure*}

Figure \ref{fig:sim} shows the results in our high-fidelity simulation environment. In particular, Figure \ref{fig:sim} (right) depicts the evolution of the DTBFs over the whole experiment. As it can be seen, for the nominal policy, the Segway fails to satisfy the mission specifications~(since $h$ becomes negative in many instances). However, with the safety-shield the satisfaction of mission specifications is guaranteed ($h$ is always positive). Furthermore, the finite time DTBF becomes positive at the end of the experiment, which shows that the \emph{eventually} specification in~\eqref{eq:specexample} is satisfied. More information on this simulation can be found in the video here \cite{youtubeclip}.

\section{Conclusions} \label{sec:conclusions}

We proposed a technique based on DTBFs to enforce safety/mission specifications in terms of LDTL formulas. We elucidated the efficacy of our methodology with simulations involving a heterogeneous robot team. Future research will explore  policy synthesis for POMDPs ensuring both safety and optimality. To this end, we use receding horizon control with imperfect information~\cite{5509278} and control invariant set estimation of MPOMDPs via Lyapunov functions~\cite{ahmadi2019control}. More general mission specifications can also be studied using time-varying DTBF (see~\cite{8404080} for time-varying barrier functions for continuous dynamics subject to STL specifications).

\footnotesize{
\bibliography{references}
}
\bibliographystyle{plain}

\end{document}